\documentclass[article, twocolumn,
 amsmath,amssymb,
 aps,
]{revtex4}

\usepackage{graphicx}
\usepackage{dcolumn}
\usepackage{bm}
\usepackage[table,xcdraw]{xcolor}

\usepackage{tikz}
\definecolor{fgreen}{RGB}{204,223,181}
\definecolor{arm}{RGB}{100,140,171}
\usetikzlibrary{arrows.meta, fit, shapes, positioning}

\begin{document}


\title{Entropy corrected geometric Brownian motion}


\author{Rishabh Gupta$^{1}$}
\author{Ewa A. Drzazga-Szcz{\c{e}}{\'s}niak$^{2}$}
\author{Sabre Kais$^{1, 3}$}
\author{Dominik Szcz{\c{e}}{\'s}niak$^{4}$}
\email{d.szczesniak@ujd.edu.pl}


\affiliation{
${^1}$Department of Chemistry, Purdue University, West Lafayette, Indiana 47907, United States,\\
${^2}$Department of Physics, Faculty of Production Engineering and Materials Technology, \mbox{Cz{\c{e}}stochowa University of Technology,} 19 Armii Krajowej Ave., 42200 Cz{\c{e}}stochowa, Poland,\\
${^3}$ Department of Physics and Astronomy, and Purdue Quantum Science and Engineering Institute, Purdue University, West Lafayette, Indiana 47907, United States, \\ 
${^4}$Institute of Physics, Faculty of Science and Technology, Jan D{\l}ugosz University in Cz{\c{e}}stochowa, 13/15 Armii Krajowej Ave., 42200 Cz{\c{e}}stochowa, Poland
}

\date{\today}


\begin{abstract}
 
The geometric Brownian motion (GBM) is widely employed for modeling stochastic processes, yet its solutions are characterized by the log-normal distribution. This comprises predictive capabilities of GBM mainly in terms of forecasting applications. Here, entropy corrections to GBM are proposed to go beyond log-normality restrictions and better account for intricacies of real systems. It is shown that GBM solutions can be effectively refined by arguing that entropy is reduced when deterministic content of considered data increases. Notable improvements over conventional GBM are observed for several cases of non-log-normal distributions, ranging from a dice roll experiment to real world data.

\end{abstract}

\maketitle

\section{Introduction}

The geometric Brownian motion (GBM) has long served as a foundational model for capturing stochastic nature of systems characterized by the continuous random fluctuations. In particular, this model has has been argued to be well suited for forecasting diffusion processes \cite{grebenkov2021}, population dynamics \cite{stojkoski2019} or most notably stock prices \cite{black1973}. However, central to GBM is the assumption that logarithm of its solutions result in a normal distribution, a premise that, while convenient for analytical purposes, may not fully encapsulate the complexities inherent in realistic systems \cite{marathe2005}. For instance, real data distributions often exhibit characteristics such as non-zero skewness, excess kurtosis or fluctuating volatility, notably deviating from the idealized bell curve \cite{stojkoski2020}. These deviations have significant implications for the accuracy of traditional GBM, by challenging its abilities to properly portray extreme events or interpret underlying dynamics of considered data. 

By recognizing these limitations, a motivation arises for the exploration of solutions that go beyond the described constraints of normality. While alternative models, such as the stochastic volatility \cite{heston1993} and the jump-diffusion process \cite{merton1976}, have been proposed, they mostly call for the modification or replacement of GBM. Hence, the existing advantages of this framework, in terms of its simplicity, efficiency and continuity, are somewhat hindered or lost. As an alternative, the present study proposes {\it non-violent} approach of employing entropy constraints to improve predictive capabilities of GBM without sacrificing its beneficial features. The idea originates from the Shannon's information theory, where entropy plays pivotal role and stands for the minimum number of logical states needed to communicate some message \cite{shannon1948}. This is to say, the well-ordered message is expected to have a lower entropy than a more random one. As a result, it can be argued that entropy provides a means to judge how well GBM predicts future events by measuring the level of uncertainty around forecast data with respect to the original time series. This premise is additionally reinforced when invoking the ability of entropy measure to capture extreme events \cite{szczesniak2023}.

To better picture above reasoning let us conceive {\it gedanken} experiment that hinges on a dice roll. We begin with a well known fact that probability of getting any side number during a conventional dice roll is equally probable, resulting in a uniform distribution. According to Shannon, the information about such process can be quantified by the entropy given as \cite{shannon1948}:
\begin{eqnarray}
H=-\sum_{i=1}^{N}p_{i}\ln{p_{i}},
\end{eqnarray}
where $N$ is the number of intervals or classes in the probability distribution (PD) and $p_{i}$ denotes the corresponding probability. For the mentioned conventional dice roll, the Shannon entropy takes its {\it maximum} value of $2.585$, as $p_{i}=1/6$. However, if there is a biased dice roll with the probability of getting some particular number equal to 1/2 and the probability of 1/10 corresponding to each of the remaining outcomes, we have a more deterministic scenario where $H = 2.161$. Importantly, further increase in the probability of rolling particular number leads to the continued decrease in entropy. Henceforth, it can be established that an increase in the deterministic content within a probability distribution corresponds to a reduction in entropy.

Extending the described principle to the domain of evolving time series reveals that normality assumption may indeed be considered as a rough approximation, particularly within a short to mid-term time frame. Consequently, employing GBM for this purpose should prove less efficacious. The observed correlation between an augmented determinism level and decreased system entropy reinforces presented viewpoint. In alignment with this rationale, an exploration of the entropy change, resulting from the addition of a trajectory forecast by GBM to the original time series, is warranted. In particular, an analysis of such entropy shift should allow for the identification of trajectories that amplify the preeminent probabilities within the original time series, thereby leading to a reduction in entropy. Here, a comprehensive analysis, underlined by this conceptual framework, is provided including not only introduction of the corresponding methodology but also its detailed validation.

\section{Methodology} \label{method}

To begin with, it is instructive to recall and briefly discuss the GBM approach. This stochastic process constitutes extension of conventional Brownian motion by incorporating drift and volatility terms. Its mathematical derivation involves solving the following stochastic differential equation:
\begin{eqnarray}
    dS(t) = \mu S(t)dt + \sigma S(t) dW(t),
\end{eqnarray}
where $S(t)$ is the series value at time $t$, $\mu$ and $\sigma$ are respectively the drift and volatility coefficients, while $dW(t)$ denotes the Wiener process (the Brownian motion). By using It\^{o}'s calculus and assuming log-normal distribution of considered data under GBM, the central equation of $S(t)$ can be obtained \cite{ito1944,ito1951a,ito1951b}:
\begin{eqnarray}
    S(t) = S(0)\exp((\mu-\sigma^2/2)t + \sigma W(t)).
    \label{gbm}
\end{eqnarray}
In this respect, GBM is a memoryless process, meaning its future behavior does not depend on past states. Because of the stochasticity imparted by the Wiener process, GBM allows to forecast numerous trajectories using $\mu$ and $\sigma$ of the corresponding probability distribution. This yields solutions that can be next employed to compute expectation value of the averaged future movement.

Here, the above approach is modified according to the presented concept of entropy constraints. This modification requires that entropy calculated for a given distribution is compared with a reference state. Thus, higher the entropy with respect to the reference state, higher will be the randomness. Following reverse analogy, the entropy would decrease if the level of determinism increases within the probability distribution. In this context, the main aim of the proposed modification is to search for the series predictions that lead to the described shift in entropy. This would ensure that, conditioned on a given probability distribution, the full information content is exploited to come up with a series of predictors for the time series that decreases the entropy of the evolved distribution. 

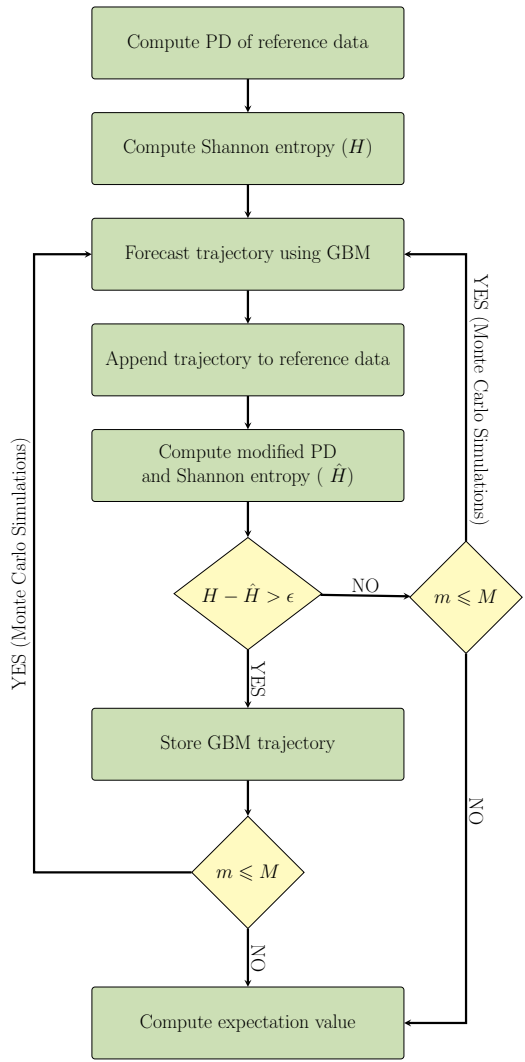
\begin{figure}[ht!]
\tikzset{
agent/.style = {draw, rounded corners, fill=fgreen!, text width=42em, minimum height=32mm, align=center, font = {\Huge}},
decision/.style={draw, diamond,aspect=2, minimum width=3cm, minimum height=1.5cm, text centered, fill=yellow!30, node distance=3.5cm, font = {\Huge\sffamily}}
}
\resizebox{!}{14cm}
{
\begin{tikzpicture}[sibling distance=100pt, node distance = 4.7cm, auto,>=stealth]
\tikzstyle{line} = [draw, thick, color=black, -latex', line width=0.7mm]
 
\node[agent] (A){Compute PD of reference data};
\node[agent, below of=A, node distance=4.7cm] (B){Compute Shannon entropy ($H$)};
\node[agent, below of=B , node distance=4.7cm] (C){Forecast trajectory using GBM};
\node[agent,below of=C, node distance=4.7cm] (D){Append trajectory to reference data};
\node[agent,below of=D, node distance=4.7cm] (E){Compute modified PD and Shannon entropy ( $\hat{H}$)};
\node[decision, below of=E, node distance=5.7cm, minimum size=5cm] (F){ $H-\hat{H}>\epsilon$};
\node[decision, below right=7.7cm and 1.5 cm of D, minimum size=5cm] (G){ $m \leqslant M$};
\node[agent, below of=F, node distance=6.7cm] (I){Store GBM trajectory};
\node[decision, below of=I, node distance=5.7cm, minimum size=5cm] (J){ $m \leqslant M$};
\node[agent, below of=J, node distance=6.7cm] (H){Compute expectation value};

\draw[->,thick, line width=1mm] (A) -- (B);
\draw[->,thick, line width=1mm] (B) -- (C);
\draw[->,thick, line width=1mm] (C) -- (D);
\draw[->,thick, line width=1mm] (D) -- (E);
\draw[->,thick, line width=1mm] (E) -- (F);
\draw[->,thick, line width=1mm] (G.north) -- node[midway, sloped, above, rotate=180, font = {\Huge}]{YES (Monte Carlo Simulations)} ++(0,12.5) |- (C.east);  
\draw[->,thick, line width=1mm] (F) -- node[midway,sloped,above, pos=0.5, font = {\Huge}]{NO} (G);
\draw[->,thick, line width=1mm] (F) -- node[midway,sloped,above, pos=0.5, font = {\Huge}]{YES} (I);
\draw[->,thick, line width=1mm] (I) -- (J);
\draw[->,thick, line width=1mm] (J.west) -- node {} ++(-7,0) -- node[midway, sloped, above, font = {\Huge}]{YES (Monte Carlo Simulations)} ++(0,27.5) -- (C.west);
\draw[->,thick, line width=1mm] (J) -- node[midway,sloped,above, pos=0.5, font = {\Huge}]{NO} (H);
\draw[->,thick, line width=1mm] (G.south) -- node[midway, sloped, above, font = {\Huge}]{NO} ++(0,-14) |- (H.east);
\end{tikzpicture}
}
\caption{Flowchart representing steps involved in the EC-GBM approach for time series forecasting. The Monte Carlo sub-procedure is marked and is meant to run $M$ times in total, with each iteration denoted by $m$. The EC-GBM solutions are refined with respect to the entropy threshold value $\epsilon$.}
\label{fig01}
\end{figure}

The detailed flowchart comprising steps of the proposed entropy corrected geometric Brownian motion (EC-GBM) approach is shown in Figure \ref{fig01}. In summary, GBM is employed there to generate trajectories based on the reference PD, utilizing its mean and standard deviation. Next, these trajectories are incrementally appended to this distribution, and the entropy for each new distribution is computed. A comparison is then made between the calculated entropy and the entropy of the reference distribution. If the entropy decrease surpasses a predetermined threshold, denoted as $\epsilon$, the corresponding trajectory is accepted as a forecast trajectory. The selection of $\epsilon$ can be configured as a percentage relative to the maximum observed decrease in entropy throughout the simulation. This process is iterated in a Monte Carlo fashion, yielding a set of forecast trajectories aligned with the characteristics of the underlying distribution.

\begin{figure}[h!]
\includegraphics[width=\columnwidth]{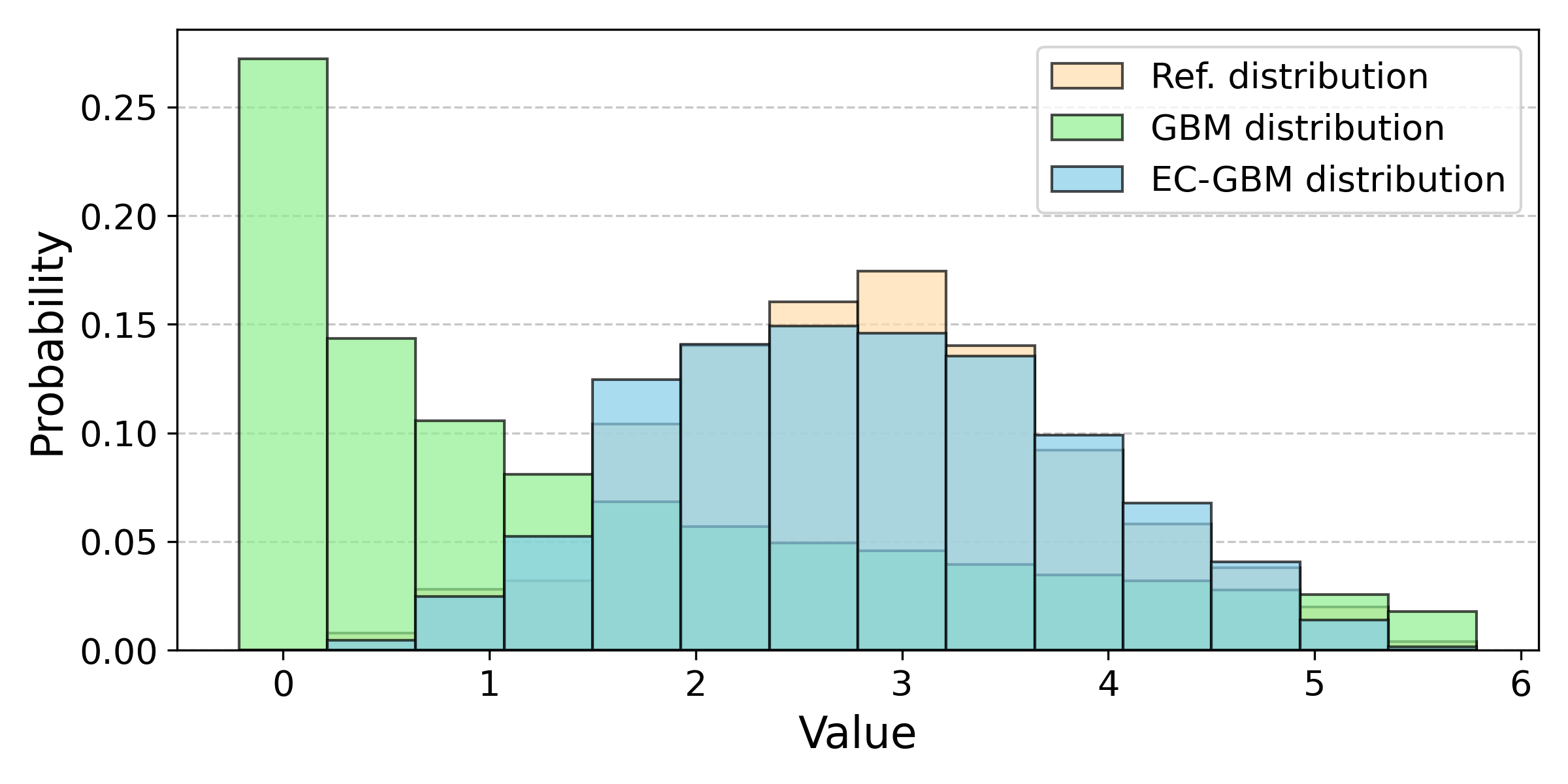}
\caption{The reconstructed probability distributions resulting from the GBM (green) and EC-GBM (blue) approaches when the underlying reference distribution is normal (orange).} 
\label{fig02}
\end{figure}

To briefly illustrate the comparative advantages of EC-GBM over conventional GBM, consider scenario where the data follows normal distribution instead of a log-normal one. The GBM approach in such circumstances would yield trajectories significantly divergent from the reference distribution {\it i.e.} GBM would assume that data adhere to a log-normal distribution, guiding its trajectory forecasts accordingly. However, as depicted in Figure \ref{fig02}, even when the corresponding distribution deviates from the log-normal character, the EC-GBM method adeptly filters trajectories in alignment with the reference distribution. This underscores the effectiveness and robustness of EC-GBM as a superior alternative to GBM, irrespective of the underlying distributional assumptions.

\section{Biased dice roll}

\begin{figure}[h!]
\includegraphics[width=\columnwidth]{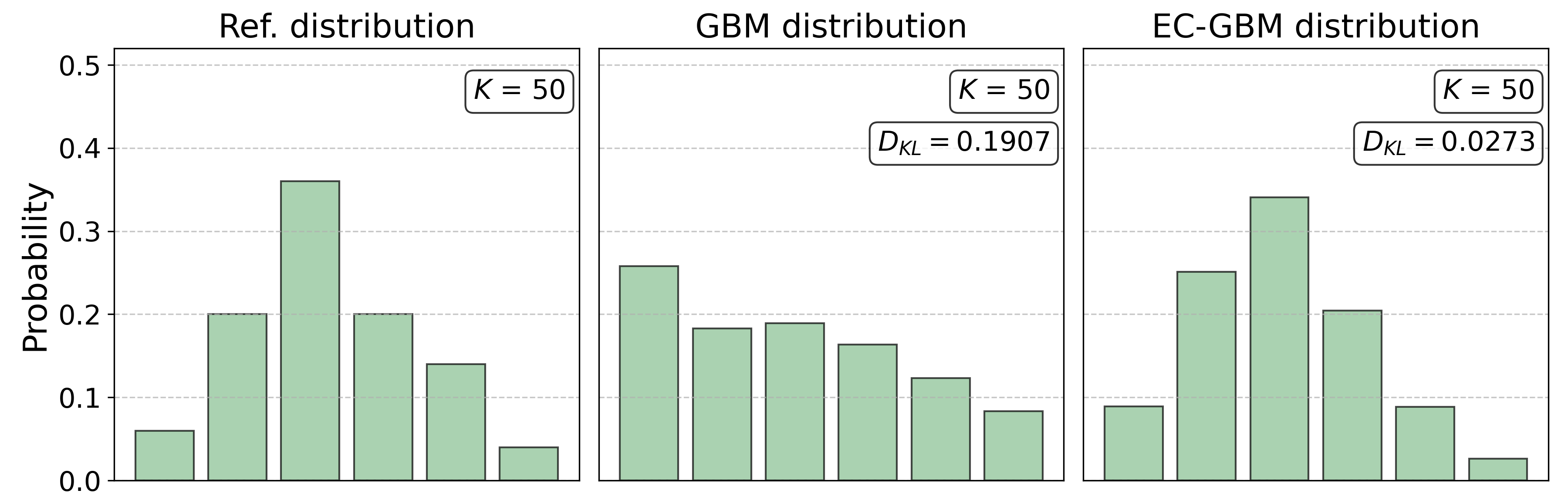}
\includegraphics[width=\columnwidth]{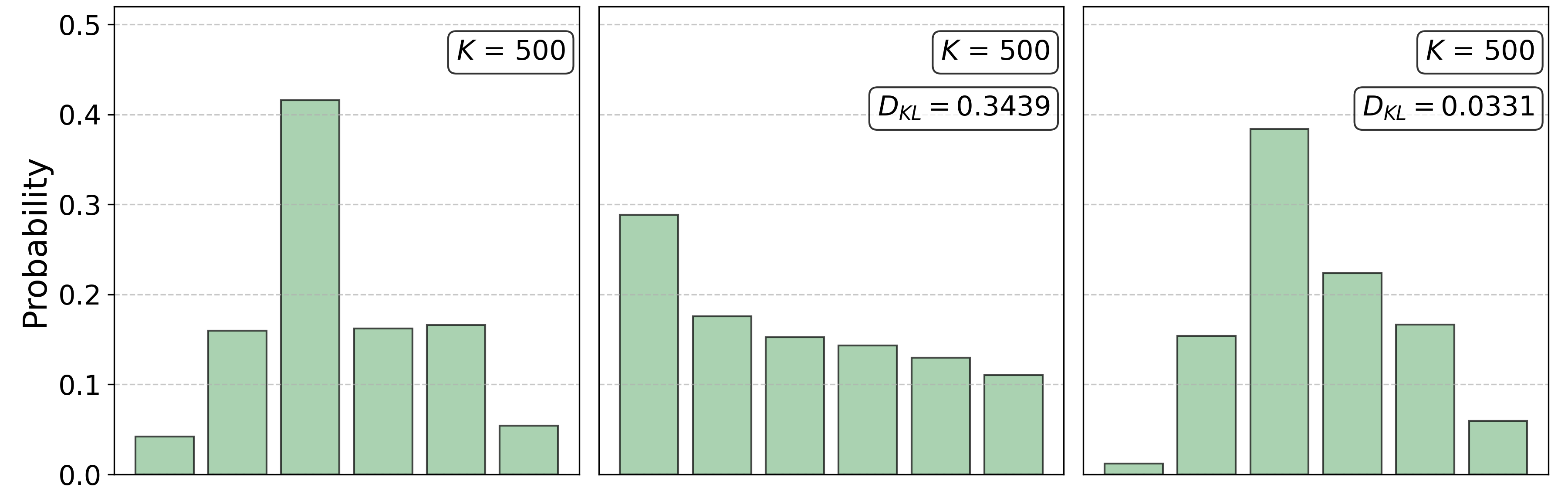}
\includegraphics[width=\columnwidth]{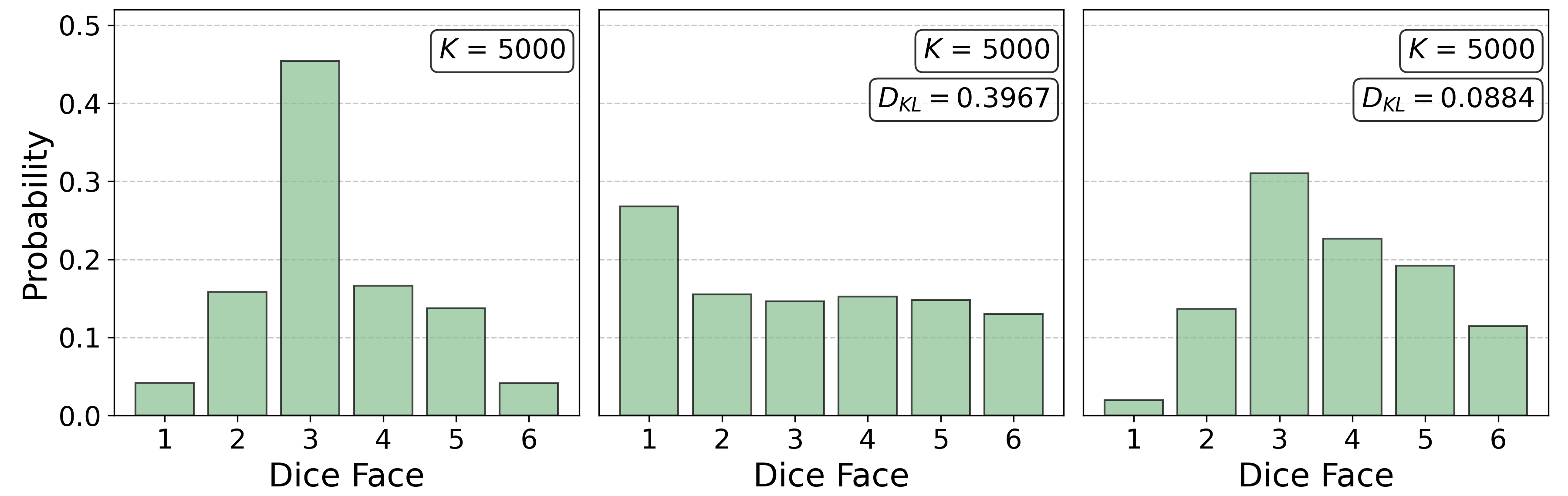}
\caption{The probability distributions of a biased dice for varying numbers of rolls, as generated through the GBM and EC-GBM approaches. This process involves considering reference probability distribution of rolls for up to $K$ rolls. The mean and the standard deviation are utilized to forecast the next 10, 50, 100 dice rolls for $K = 50$, $K = 500$, $K = 5000$ respectively, using the GBM approach. Subsequently, the trajectories that lead to decrease in the entropy upon appending to the reference distribution constitute the EC-GBM distribution.} 
\label{fig03}
\end{figure}

To properly validate the depicted potential and accuracy of EC-GBM, we analyze this approach at the quantitative level. This allows systematic and in-depth statistical discussion of the EC-GBM as well as the initial identification of the regimes where the proposed model works best. To do so, let us consider again a biased dice, however, this time the probability of rolling every face is unequal. The manifestation of such bias becomes particularly apparent when examining reference distributions depicted in Figure \ref{fig03} over an increasing number of rolls, $K\in \left< 50, 5000 \right>$. By envisaging the biased dice roll as a time series, our objective is to discern the inherent distribution and generate future rolls that are in alignment with the underlying probability distribution. In this context, it is crucial to note that the resulting distribution does not adhere to a normal distribution, even with a large number of rolls. The bias introduces skewness towards favored outcomes, deviating from the typical bell curve associated with the normal distribution.

The future trajectories are obtained by employing both the GBM and EC-GBM techniques for comparison and the resulting distributions are shown in Figure \ref{fig03}. The inadequacy of the traditional GBM approach is evident in Figure \ref{fig03} when confronted with a skewed reference distribution. It is at this juncture that the EC-GBM emerges again as a more robust method, capitalizing on the maximum information content within the time series. This allows to predict outcomes with heightened accuracy that align with the underlying probability distribution, even in scenarios with fewer rolls. In what follows, the obtained results clearly demonstrates limitations of the GBM as caused by its inherent log-normality assumption.

As a measure of the difference between the reference and generated probability distributions, given by GBM and EC-GBM methods, the Kullback-Leibler ($D_{KL}$) divergence is calculated and shown in Figure \ref{fig03}. The $D_{KL}$ measure is often used in information theory and statistics to quantify the difference between two probability distributions in the following manner \cite{kullback1951}: 
\begin{eqnarray}
D_{KL}(P||Q) = \sum_i P(i) \log(P(i)/Q(i)),
\end{eqnarray}
where \textit{P} and \textit{Q} stands for the reference and approximated distributions, respectively. This is to say, the $D_{KL}$ measures the information lost when \textit{Q} is used to approximate \textit{P}.

As depicted in Figure \ref{fig03}, the $D_{KL}$ between the reference distribution and the distribution generated using EC-GBM approach is significantly lower compared to that observed with the use of traditional GBM approach. This is due to the fact that the biasness increases the level of determinism within the probability distribution and consequently, as discussed in Section \ref{method}, it decreases the entropy of the appended trajectory that enhances the probability peaks with regards to the reference distribution. Therefore, the trajectories predicted by GBM, that contribute to a decrease in entropy, essentially signify paths reinforcing bias within the distribution. 

\section{Results}

The above discussion is supplemented by the analysis of the proposed methodology against real world empirical examples to further substantiate effectiveness of EC-GBM. Since our approach focuses on optimizing the utilization of information embedded in historical time series, it makes a perfect tool for refining forecasting models and strategies to enhance our ability to predict future data behavior. Among others, this underlies the predictive analytics in financial markets that constitutes convenient case for our considerations. In particular, we present two illustrative and interdependent examples of time series forecasting: (i) we examine two simulated time series, one characterized by an upward trend and the other by a downward trend, (ii) we apply our approach to a real-world time series obtained from the financial market for the pair made up of gold against the US dollar (for the timeframe from 9 to 17 March 2023).

\begin{figure}[h!]
\includegraphics[width=\columnwidth]{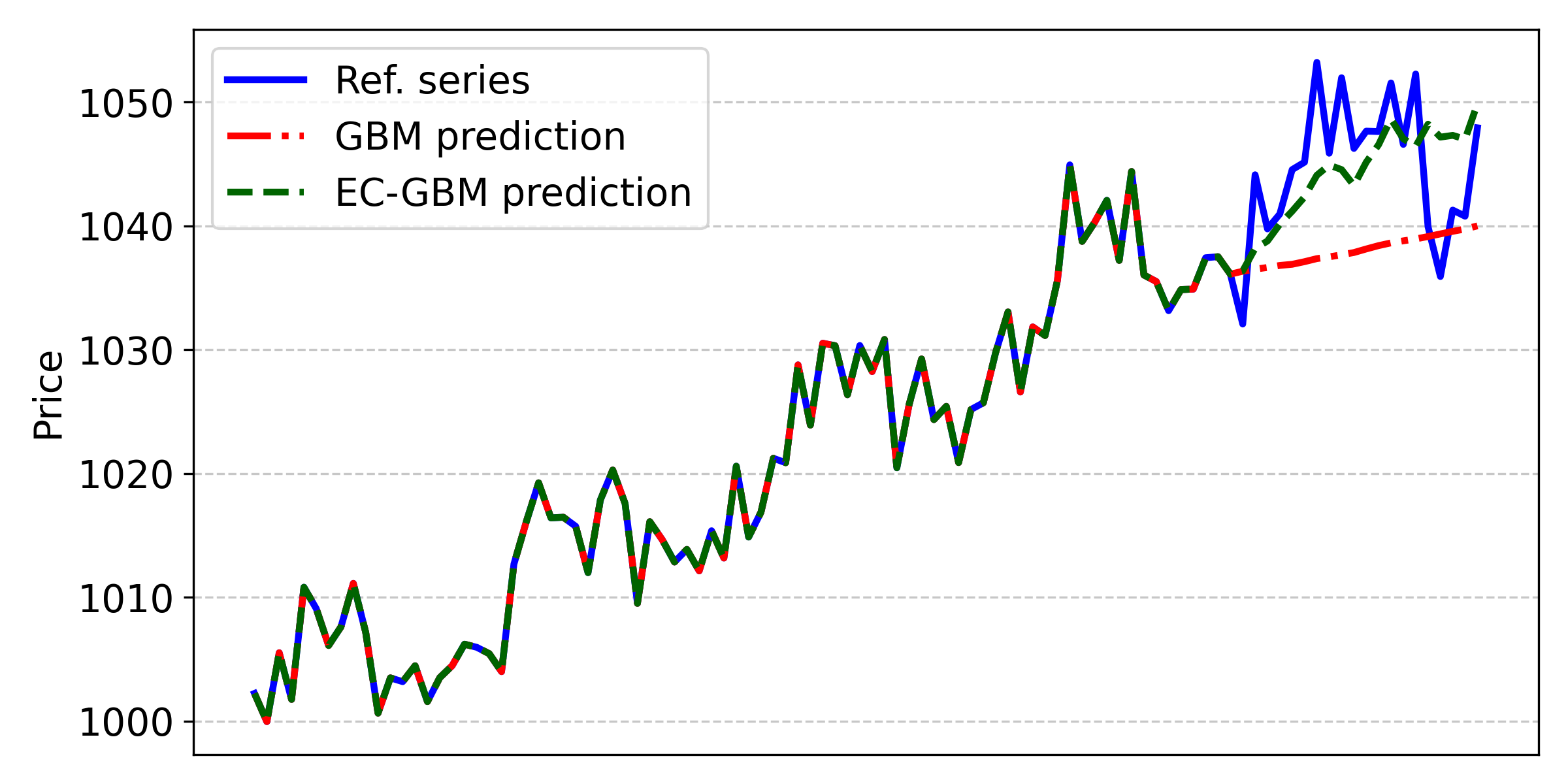}
\includegraphics[width=\columnwidth]{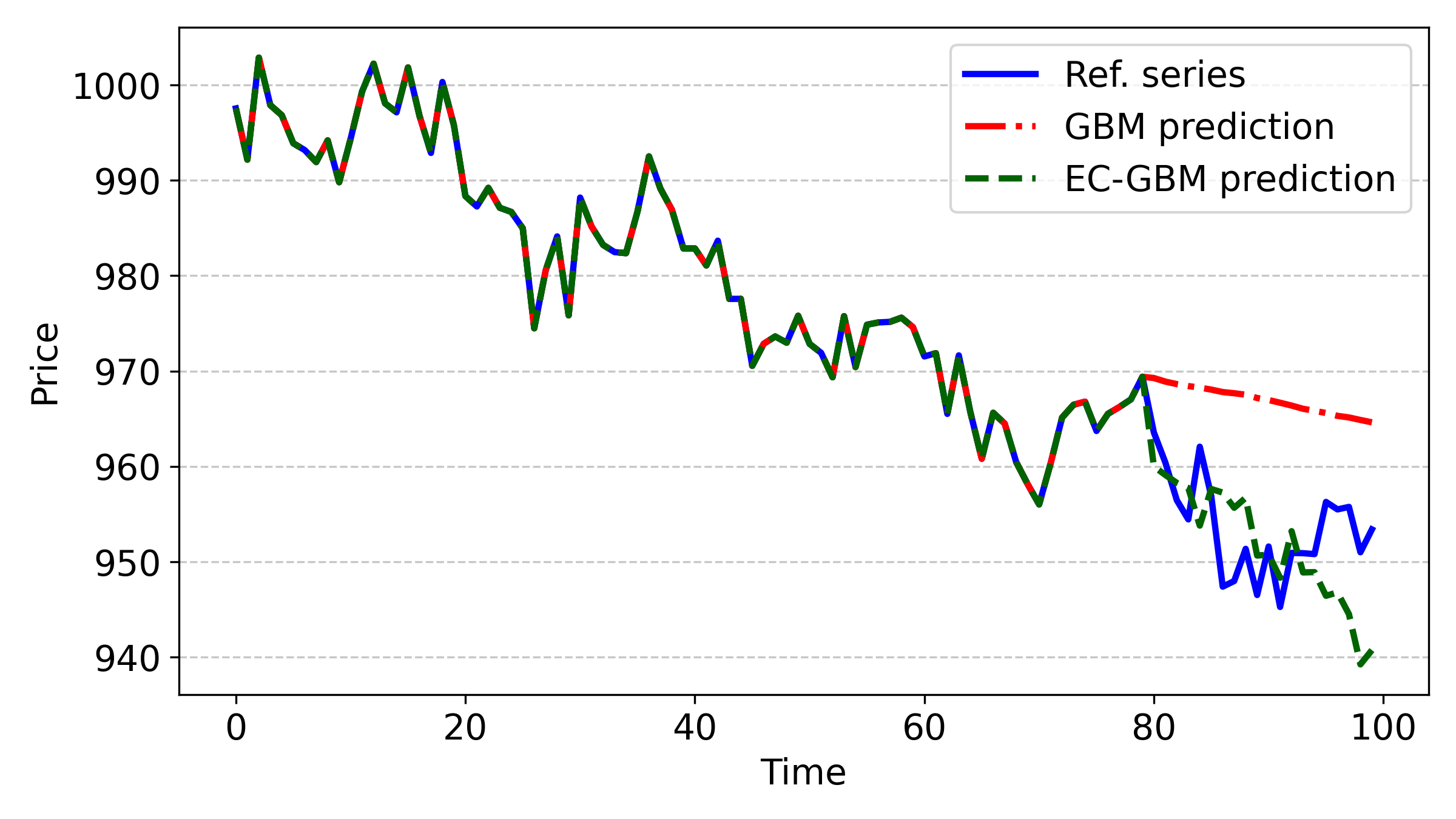}
\caption{The forecast trajectories of simulated data, as obtained within the GBM (dashed red line) and EC-GBM (dashed green line) approach. The data for the upward (top panel) and downward (bottom panel) trend is considered. For reference purposes the historical data of simulated time series is depicted (blue solid line).} 
\label{fig04}
\end{figure}

The simulated time series for the artificial upward and downward trend are respctively represented by the following equations:
\begin{eqnarray}
    X_t = X_0 + m\times(t+\sin(2t)) + \delta, \\
    X_t = X_0 - m\times(t+\sin(2t)) + \delta,
\end{eqnarray}
where $X_t$, $X_0$ are time series values at time $t$ and 0 units, $m$ is the slope and $\delta$ denotes a random error obtained from a normal distribution with mean zero and variance one. The computation of such time series is carried out for 100 steps with the objective of leveraging the initial 80 steps as an input to our methodology, subsequently employing both the GBM and EC-GBM approaches to forecast the next 20 steps. This approach allows us to evaluate the predictive capabilities of our proposed methodology in comparison to the traditional GBM method. To derive the ultimate forecast trajectory, we calculate the expectation value for both the GBM and EC-GBM approaches at each time step. In the EC-GBM approach, expectations are computed based on trajectories that exhibit a reduction in entropy of at least 75$\%$ compared to the maximum entropy reduction observed across all GBM trajectories. Figure \ref{fig04} visually demonstrates that while both GBM and EC-GBM accurately predict the overall trend, the forecast generated by the EC-GBM offers more valuable insights. Specifically, the EC-GBM solutions not only align with the temporal movement of the time series but also provide nuanced information regarding the magnitude of observed shifts, a feature that harmonizes more closely with the actual trajectory of time series. 

\begin{figure}[h!]
\includegraphics[width=\columnwidth]{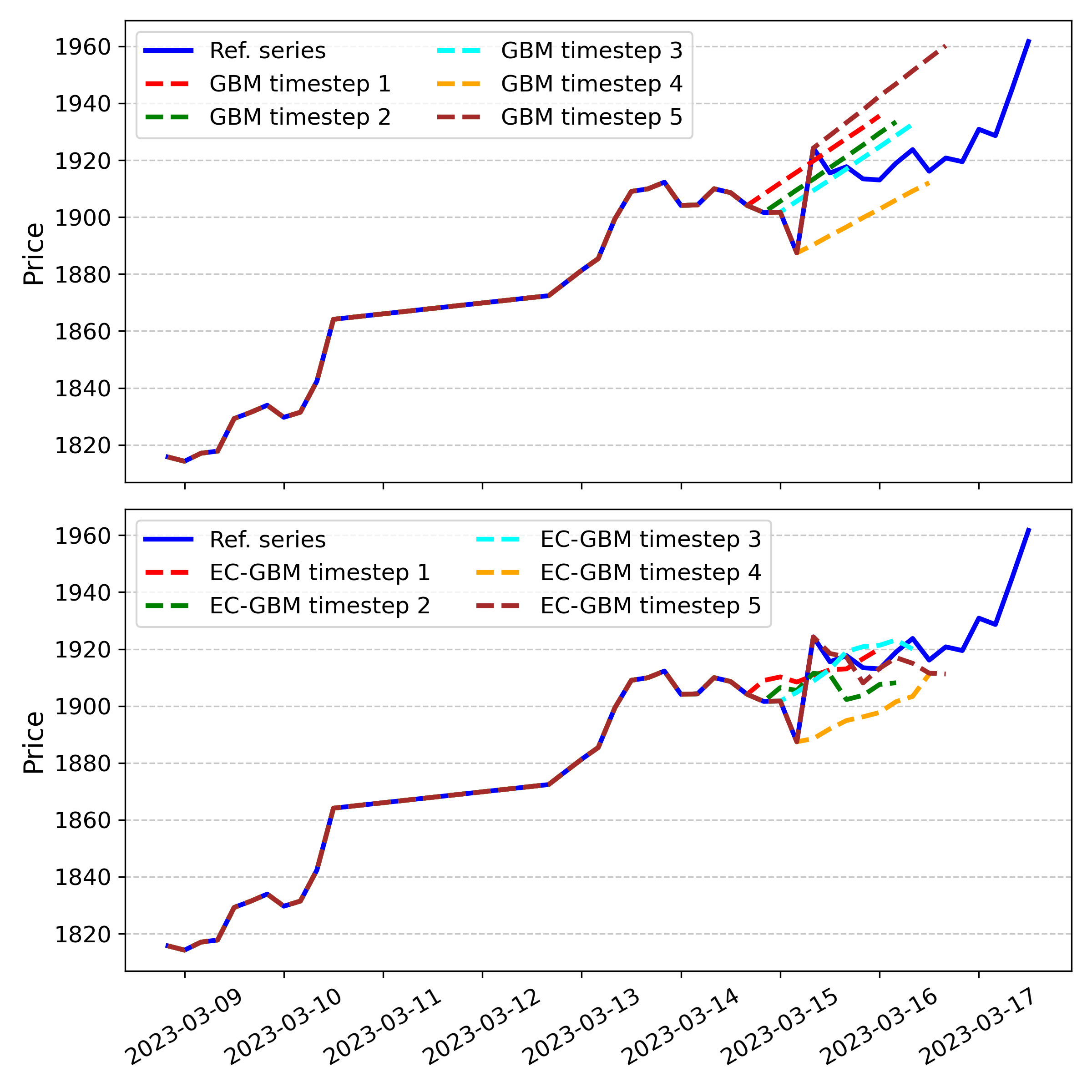}
\caption{The forecast trajectories of the real world financial data for the pair made up of gold against the US dollar (dashed lines), as obtained within the GBM (top panel) and EC-GBM (bottom panel) approach. The simulated data is presented for the selected time steps. For reference purposes the historical data of real world time series is depicted (blue solid line).} 
\label{fig05}
\end{figure}

It is crucial to note that the simulated time series under consideration incorporates both an underlying trend and seasonality, providing ample information for informed forecasting of future trajectories. In contrast, real-world time series often involve a multitude of additional factors influencing their dynamics, necessitating a more comprehensive approach to account for these complexities. Nevertheless, the incorporation of information criteria in forecasting methodologies holds promise for enhancing the precision and quality of predictions in the presence of diverse influencing factors. In this context, we also illustrate the application of our forecasting methodology on real-world financial market data as shown in Figure \ref{fig05}. To simulate real-world conditions accurately, we showcase the evolution of forecasted trajectories from both methodologies at various time steps, where each forecast utilizes the most recent 4 days of data as the underlying distribution to forecast the time series for the subsequent 32 hours, with each time step representing a 4-hour interval. Notably, both the GBM and the EC-GBM approaches adeptly predict the directional trend in market movement, similar to the previous case of the simulated time series. The correspondence is also observed when closely inspecting the EC-GBM results alone. In detail, the EC-GBM approach excels in capturing not only the qualitative trend but also the quantitative nuances of the market dynamics, offering a more comprehensive and accurate depiction of the observed movements.

\section{Conclusion}

In summary, the present research endeavor allows to deepen current understanding of how to effectively discern and harness the wealth of information contained within historical time series. It aligns with the concept that the maximization of information content within probability distributions can be well characterized by entropy, serving as a fundamental tool for quantifying uncertainty and disorder across various systems and distributions \cite{jaynes1957,rishabh2021}. In particular, the above findings are shown here to be highly advantageous for improving forecasting potential of elemental GBM process, by adopting entropy corrections to uncover patterns and structures within data sets of interest. The crux of introduced methodology lies in manipulating entropy measure through the strategic addition of trajectories, simulated by the traditional GBM method, to the existing time series distribution. By doing so, it is possible to notably refine our ability to predict and anticipate future data behavior.

Our research underscores the potential applications of the EC-GBM approach across various domains where the traditional GBM method is employed. Particularly noteworthy are its implications in risk management, such as estimating value at risk, where the inadequacies of GBM become apparent when faced with the non-log-normal distributions. Additionally, in option pricing, where GBM-based Monte Carlo simulations are prevalent, our method offers promising avenues for improvement. By addressing the limitations of GBM across these critical financial applications, our research opens doors to more robust risk assessment and pricing strategies, thereby facilitating more informed decision-making in financial markets. Furthermore, by surpassing the log-normality assumption of the underlying distribution in the EC-GBM approach, even  short-term time series distributions can be utilized to construct risk strategies tailored to the high volatility prevalent in financial markets. Nonetheless, the perspectives of the presented model are not limited only to the financial applications. Given appropriate conditions, the reasoning outlined here can be adapted to diverse fields where stochastic processes play a pivotal role. For instance, in modeling quantum stochastic processes, especially within continuous-time quantum dynamics, decoherence phenomena can be effectively described using GBM to characterize the random fluctuations responsible for decoherence. Moreover, refining probability distributions based on entropy values, as demonstrated here, shows promise for addressing analogous challenges in refining sampling schemes, including rejection sampling, importance sampling, Metropolis-Hastings algorithm, and others.

\begin{acknowledgments}
S.K. would like to acknowledge the National Science Foundation under award number 1955907.
\end{acknowledgments}

\bibliographystyle{apsrev}
\bibliography{bibliography}

\end{document}